\documentclass[%
 reprint,
showpacs,%preprintnumbers,
 amsmath,amssymb,
 aps,
prl,
]{revtex4-1}
\usepackage{epsfig}
\usepackage{epstopdf}
\usepackage{graphicx}% Include figure files
\usepackage{dcolumn}% Align table columns on decimal point
\usepackage{bm}% bold math
\begin{document}

%\preprint{APS/123-QED}
\setlength{\belowdisplayskip}{0pt} \setlength{\belowdisplayshortskip}{0pt}
\setlength{\abovedisplayskip}{0pt} \setlength{\abovedisplayshortskip}{0pt}
\title{Nonlocal cancellation of dispersion in Franson interferometry}

\author{Tian Zhong}
%\email{tzhong@mit.edu}
\author{Franco N. C. Wong}%

\affiliation{Research Laboratory of Electronics, Massachusetts Institute of Technology, 77 Massachusetts Avenue, Cambridge, Massachusetts 02139, USA}

\date{\today}% It is always \today, today,

\begin{abstract}
Dispersion and its cancellation in entanglement-based nonlocal quantum measurements are of fundamental and practical interests. We report the first demonstration of cancellation of femtosecond-level dispersion by inverting the sign of the differential dispersion between the long and short paths in only one arm of a fiber-based Franson interferometer. We restore the otherwise limited quantum visibility to an unprecedented 99.6\%, and put time-energy entanglement at the same level of quality as polarization entanglement for use in quantum information processing applications.
%Temporal dispersion and its cancellation are critical issues in entanglement-based nonlocal measurements. Our experiment sheds light on nonlocal application of dispersion cancellation to fundamentally quantum measurements with no classical analog.

\end{abstract}
\pacs{03.65.Ud, 03.67.Mn, 42.65.Lm}% PACS, the Physics and Astronomy
\maketitle

%\tableofcontents

Nonlocality is the quintessential quantum property of entanglement, in which a local action made in one subsystem can influence the other subsystem at a remote location in a way that cannot be explained by classical or local hidden-variable theories. The most well known nonlocality test is the violation of Bell's inequality such as its Clauser-Horn-Shimony-Holt (CHSH) form for polarization-entangled photons \cite{CHSH} or Franson interferometry for two time-energy entangled photons \cite{Franson1989}. 

In virtually all nonlocal experiments, photons could be broadened traveling through dispersive media unless the nonlocal measurements are immune to dispersion or compensation is administered. Polarization measurements in the CHSH form of Bell's inequality violation are not sensitive to temporal spread due to dispersion, and this is corroborated by excellent nonlocal measurements of entanglement quality made without any attention paid to dispersion, with two-photon quantum-interference visibility of $>$99\% \cite{kwiat, kim}. On the other hand, Franson interferometry is expected to be sensitive to dispersion and the best measurement to date \cite{Zhong} without any dispersion consideration does not quite measure up to that of polarization entanglement. It is therefore of fundamental interest to ponder the question of applying dispersion compensation nonlocally and how that may affect nonlocal measurements of entanglement quality. That is, can dispersion at one location be compensated by action at a different location? In this Letter, we investigate the effects of dispersion in Franson interferometry and demonstrate for the first time complete nonlocal cancellation of femtosecond-level dispersion in classicality-violating nonlocal quantum measurements.

It is instructive to briefly examine some common dispersion cancellation techniques and applications. It is well known that Hong-Ou-Mandel (HOM) interferometry is immune to even-order dispersion \cite{Steinberg}, and this immunity is exploited in various applications such as high-precision clock synchronization \cite{Giovannetti} and resolution-enhanced quantum optical coherence tomography (OCT) \cite{qoct}. However, HOM interferometry is a local interference measurement of two incident photons that do not need to be entangled. Indeed, classical interference signature analogous to the HOM interference dip has been observed in a quantum-mimetic experiment using oppositely chirped laser pulses \cite{Resch}, and dispersion cancellation has been demonstrated in phase-conjugate OCT using classical phase-sensitive cross-correlated Gaussian light \cite{pc-oct}. Franson proposed a truly nonlocal dispersion cancellation experiment in which the joint temporal correlation between two time-energy entangled photons would remain unchanged after they have separately gone through media with opposite signs of dispersion \cite{Franson92}. However, such dispersion cancellation on typically sub-picosecond temporal correlation is difficult to verify experimentally because it requires detection timing accuracy at the femtosecond level. Two recent experiments hint at the correctness of nonlocal dispersion cancellation. O'Donnell recombined the two photons after dispersion and utilized time-resolved upconversion to observe dispersion cancellation at the femtosecond level \cite{Donnell}; however, the observation was based on local measurements. Baek \emph{et al}. used strong dispersion to broaden the effective pulse duration of the entangled photons and observed narrowing that was consistent with nonlocal dispersion cancellation \cite{Kim}. However, their experiment was detector resolution limited and could not demonstrate narrowing to the original sub-ps temporal correlation. 

In this work, we choose Franson interferometry to demonstrate the essential physics of Franson's nonlocal dispersion cancellation scheme. The interferometric scheme allows us to characterize dispersion at the femtosecond level without detector resolution limitation, and to study the effect in a nonlocal quantum measurement setting that requires entanglement and violates classicality. We show that only the differential dispersion within the unbalanced paths of each fiber-based Mach-Zehnder interferometer (MZI) affects the Franson interference, which is not sensitive to the dispersion between the source and the two MZIs. In addition, from a practical point of view, dispersion compensation in Franson interferometric measurements allows its quantum-interference visibility to reach 99.6\% that is on par with the best CHSH measurements obtained with polarization-entangled photons, which bodes well for applications that utilize time-energy entanglement.  

Figure~1 shows a model of dispersive Franson interferometry. Continuous-wave (cw) spontaneous parametric downconversion (SPDC) generates time-energy entangled signal and idler photons, which propagate independently to two separate locations where a Franson interferometric measurement is performed nonlocally. Each arm of the Franson interferometer consists of an all-fiber unbalanced MZI using 50:50 fiber beam splitters. We note that only two types of dispersion are relevant to our setup. The first is the dispersion along the path  from the SPDC source to the MZIs, and the second is the differential dispersion between the long and short paths in each MZI. The former does not have any impact on the Franson interference visibility, the reason of which will become clear later in the analysis. Therefore, in our model only the differential dispersion is included, as denoted by unitary dispersion operators $\hat{D}_1$, $\hat{D}_2$. 
\begin{figure}[hbt]
\includegraphics[width=0.48\textwidth]{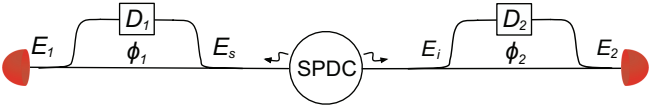}% Here is how to import EPS art
\caption{\label{f1} Model of fiber-based Franson interferometric measurements including differential dispersion $D_1$ and $D_2$.}
\end{figure}

The electric field operators before and after the Franson interferometer are denoted by $\hat{E}_{s,i}(t)$ and $\hat{E}_{1,2}(t)$, respectively, with
\begin{equation}
\hat{E}_{s,i}(t)=\int d\omega_{s,i}e^{-i\omega_{s,i} t}\hat{a}_{\omega_{s,i}}\,,
\end{equation} 
where $\hat{a}_{\omega_{s,i}}$ is the photon annihilation operator for frequency  $\omega_{s,i}$. In the Heisenberg representation, the unitary transformations of $\hat{D}_{1,2}$ can be easily incorporated by writing the field operator at the detectors as
\begin{equation}
\hat{E}_{1,2}(t)=\frac{1}{\sqrt{2}}(\hat{E}_{s,i}(t)+e^{i\phi_{1,2}}\hat{D}_{1,2}^\dagger\hat{E}_{s,i}(t-\Delta T)\hat{D}_{1,2})\,,
\end{equation}
where we have ignored the common delay through the MZI's long and short paths, and $\Delta T$ is the propagation time difference between the long and short paths, which is assumed to be the same for both MZIs and much larger than the biphoton correlation time. $\phi_{1,2}$ are two independent phase controls. The operators in the second term can be conveniently evaluated in the frequency domain,
\begin{eqnarray}
\hat{D}^\dagger_{1,2}\hat{E}_{s,i}&&(t-\Delta T)\hat{D}_{1,2}= \nonumber\\
&&\int d\omega_{s,i}e^{-i\omega_{s,i}(t-\Delta T)} e^{-i\Phi_{s,i}(\Omega_{s,i})}\hat{a}_{\omega_{s,i}}\,,
\end{eqnarray}
where $\Omega_{s,i}=\omega_{s,i}-\omega_p/2$ is the frequency detuning from the central wavelength which is taken as $\omega_p/2$ for simplifying the analysis. The differential phase delay $\Phi_{s,i}$ is of the form
\begin{equation}
\Phi_{s,i}(\Omega_{s,i})=\sum_{n\geq 2}\frac{\Omega^n_{s,i}}{n!}\Delta(\beta_n L)\,,
\end{equation}
with $\beta_n$ being the $n^{th}$-order dispersion coefficient of the fiber, of which only the group velocity dispersion (GVD) $\beta_2$ term is dominant. Here $\Delta(\beta_n L)$ denotes the differential amount between the long and short paths. We calculate the coincidence rate $C$ between two detectors as in \cite{Franson1989}:
\begin{widetext}
\begin{eqnarray}
C&\propto &\langle\hat{E}^\dagger_1(t) \hat{E}^\dagger_2(t) \hat{E}_2(t) \hat{E}_1(t) \rangle \\
  &\propto &\langle\hat{E}^\dagger_s(t) \hat{E}^\dagger_i(t) \hat{E}_i(t) \hat{E}_s(t) \rangle + \langle\hat{D}^\dagger_1\hat{E}^\dagger_s(t-\Delta T)\hat{D}_1\hat{D}^\dagger_2\hat{E}^\dagger_i(t-\Delta T)\hat{E}_i(t-\Delta T) \hat{D}_2\hat{D}_1^\dagger\hat{E}_s(t-\Delta T) \hat{D}_1\rangle \nonumber\\
  &&+ \left\{e^{i(\phi_1+\phi_2)}\langle\hat{E}^\dagger_s(t) \hat{E}^\dagger_i(t)\hat{D}^\dagger_2\hat{E}_i(t-\Delta T) \hat{D}_2\hat{D}^\dagger_1\hat{E}_s(t-\Delta T)\hat{D}_1 \rangle+ h. c.\right\}\,.
\end{eqnarray}
\end{widetext}

In Eq.~(6) we assume that the detector timing jitters and the broadened biphoton temporal correlation are much smaller than $\Delta T$ so that the photons traveling the long and short paths remain non-coincident. By substituting (1) and (3) into Eq.~(6), we obtain
\begin{eqnarray}
C\propto\int d\omega_sd\omega_i&&\langle \cos ^2[\frac{\tilde{\phi}+\omega_p\Delta T-(\Phi_s(\Omega_s)+\Phi_i(\Omega_i))}{2}] \nonumber\\
&&\times \hat{a}^\dagger_s\hat{a}^\dagger_i\hat{a}_i\hat{a}_s\rangle.
\end{eqnarray}
The Franson interference visibility is given by $V$\,=\,$(C_{\rm max}-C_{\rm min})/(C_{\rm max}+C_{\rm min})$, where $C_{\rm max}$ and $C_{\rm min}$ are the maximum and minimum coincidences obtained by varying $\tilde{\phi}=\phi_1+\phi_2$. Note that $\Delta\omega_p\Delta T\ll1$ for a cw pump. 

In examining Eq.~(7) we note that any dispersion imposed on the biphoton prior to entering the interferometer does not affect the coincidences, since the integration does not involve relative phase between the frequency modes. Only the differential dispersion at each MZI degrades the Franson visibility, unless $\Phi_s(\Omega_s)+\Phi_i(\Omega_i)=0$. The obvious solution is to cancel the dispersion locally with $\Phi_s(\Omega_s)=\Phi_i(\Omega_i)=0$. However our primary interest and the emphasis of this letter is to achieve complete dispersion cancellation nonlocally with $\Phi_i(\Omega_i) = -\Phi_s(\Omega_s)$ and restore the Franson visibility accordingly. 

%Our interferometric setup in Fig.~\ref{f1} can be considered as adding a short dispersionless reference path between the source and each detector to the Franson's original nonlocal dispersion cancellation scheme in \cite{Franson92}. The nature of entanglement dictates a coherent superposition of the biphoton state going through the dispersive long paths and the dispersionless short paths, which yields a maximum 100\% Franson quantum-interference visibility only if the initial entanglement is perfect and the broadening of the temporal correlation due to the long dispersive paths is completely compensated. Therefore 

Our dispersion cancellation using Franson interferometry demonstrates the same underlying physics as Franson's original scheme. First, to preserve all the properties of the initial biphoton temporal correlation, quantum entanglement is necessary for complete nonlocal cancellation \cite{Wasak, Franson10}. Using classical fields with perfect frequency correlation in the original Franson scheme would at best yield a partial narrowing of the detected temporal correlation accompanied by a significant dc background \cite{Wasak, Shapiro, Franson10}. If the same frequency correlated classical fields were used in Franson interferometry, the observed visibility would be no greater than 50\% \cite{Fransonclass}. Second, the demonstrated dispersion cancellation is strictly nonlocally applied without bringing the two photons together, which echoes the essence of the Einstein-Podolsky-Rosen (EPR) paradox \cite{epr}: local action by one subsystem (dispersion at signal side) can be nonlocally influenced by a different subsystem (compensation at idler side). 

%We emphasize that the difference between our scheme and that in \cite{Franson92} is in the way the biphoton temporal correlation is measured. In \cite{Franson92}, dispersion cancellation is observed directly through narrowing of the biphoton correlation time to its initial sub-ps range, which is practically impossible given the limited timing resolution of currently available detectors. In our scheme we take advantage of the interferometric arrangements to overcome the limitations of slow detector response for measuring the biphoton temporal correlation very precisely. 

We set up the experiment shown in Fig.~2. Frequency-degenerate time-energy entangled photons at 1560\,nm were efficiently generated in a type-II single-spatial-mode PPKTP waveguide source via cw SPDC with a brightness of $10^7$ pairs/s per mW of pump \cite{Zhong}. After coupling into a polarization-maintaining fiber, the orthogonally polarized signal and idler photons were separated using a fiber polarizing beam splitter and sent to their respective arms of the Franson interferometer. The coincidence measurement was performed using two 20\% efficient self-differencing InGaAs single-photon avalanche photodiodes (SPAD) with square-wave gating at a 628.5-MHz repetition rate \cite{Zhong}. The detector timing resolution is $\sim$100\,ps, which is much greater than the picosecond biphoton correlation time. The SPADs had low after-pulse probability of \textless6\% and dark counts of \textless2$\times$10$^{-6}$ per gate. To achieve long term phase stability, the fiber interferometer was enclosed in a multi-layered thermally insulated box, whose inside temperature was actively stabilized to within 10 mK. The path mismatch $\Delta $T=4.77\,ns was set to match the duration of 3 detector-gating periods. The difference in the two path mismatches was fine tuned using an additional closed-loop temperature control of the long-path fiber in the upper arm to less than the biphoton coherence time of $\sim$1\,ps. The variable phase shift of each arm was set by a piezoelectric transducer (PZT) fiber stretcher. 
\begin{figure}[hbt]
\includegraphics[width=0.48\textwidth]{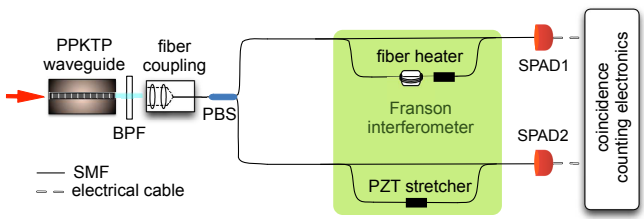}% Here is how to import EPS art
\caption{\label{f2} Experimental setup of dispersive Franson interferometric measurement. BPF: band-pass filter, PBS: polarizing beam splitter, PZT: piezoelectric transducer.}
\end{figure}

Figure~3 shows the spectra of signal and idler photons. The biphoton phase-matching bandwidth was $\sim$1.6 nm at full-width-half-maximum (FWHM). For an interferometer that was constructed using standard single-mode fibers (SMFs) with $\beta_2$\,=\,$-22.5$ fs$^2$/mm, the expected temporal spread was 26\,fs, which would only increase the $\sim$1\,ps biphoton coherence duration by less than 1\,fs if a Gaussian spectral shape is assumed. However, using Eq.~(7) to numerically integrate over the spectra in Fig.~3, we predict a 1.3\% degradation of Franson visibility due to dispersion caused by the broad spectral pedestals outside of the central lobe in Fig.~3. Therefore, to measure the intrinsic entanglement quality, such femtosecond-level dispersion should be compensated and all other sources of degradation be minimized. 

\begin{figure}
\includegraphics[width=0.5\textwidth]{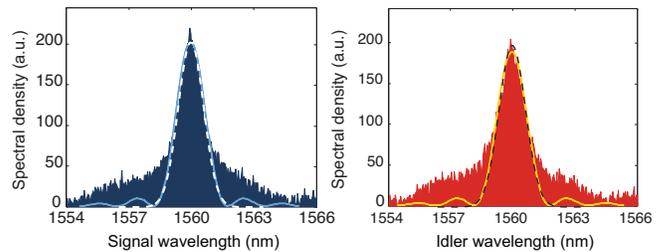}% Here is how to import EPS art
\caption{\label{f2b} Spectra of signal, idler photons showing $\sim$1.6 nm bandwidth at FWHM and noticeable spectral pedestals. Solid curves are $sinc^2$ fits. Dashed curves are Gaussian fits.}
\end{figure}

Recently we have reported \cite{Zhong} a raw Franson visibility of 98.2\,$\pm$\,0.3\% with a mean photon pair per gate $\alpha$=0.24\% for an all-SMF interferometer configuration of Fig.~4(a). Measured visibility included accidental coincidences and the 0.3\% uncertainty was based on one standard deviation of 100 measurement samples collected over a total duration of 10 s. Our high-speed data acquisition was largely the result of a reasonably good system efficiency of the source-detector combination. We noted in Ref.~\cite{Zhong} that a visibility degradation of $\sim$0.4\% was due to multi-pair events and accidentals, and the remaining 1.4\% degradation was attributed to dispersion (but without supporting evidence). Here we verified fiber dispersion was the origin of the degradation by applying narrowband spectral filtering to the biphoton, as shown in Fig.~4(b). With the same $\alpha$, we measured a raw visibility of 99.4\,$\pm$\,0.3\% after 0.36-nm bandpass filtering of the biphoton spectra. We should point out the extra loss incurred by narrowband filtering and the resultant reduction of pair flux. 
\begin{figure}[hbt]
\includegraphics[width=0.48\textwidth]{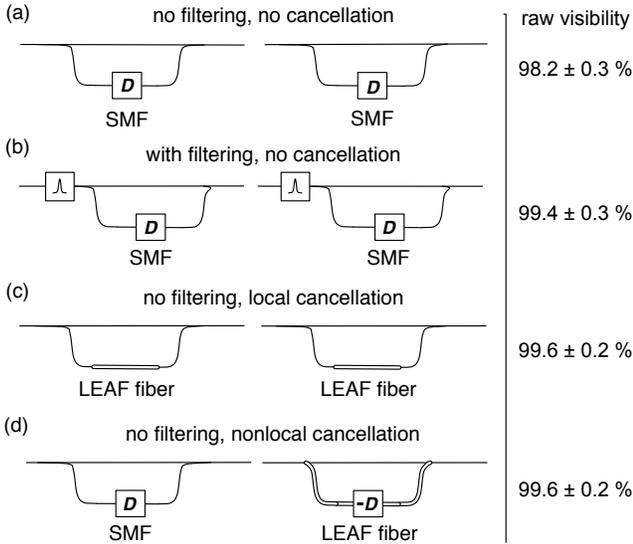}% Here is how to import EPS art
\caption{\label{f4} Configurations of Franson interferometer for testing dispersion effects. The respective measured raw visibility is shown to the right of each configuration.}
\end{figure}

A better way to restore perfect Franson interference is to cancel dispersion of the fiber interferometer without restricting the biphoton bandwidth. In Fig.~4(c), for benchmarking, we implemented an interferometer with its dispersion canceled locally by replacing a portion of the long-path SMF with low-dispersion LEAF fiber ($\beta_2$\,=\,$-6.19$ fs$^2$/mm at 1560 nm) such that the differential dispersion $\Phi_{s,i}$ was zero. We chose LEAF fiber because it has a core dimension very close to SMF so that splicing loss would be minimum. Without loss of flux, we measured a raw visibility of 99.6\,$\pm$\,0.2\% at the same $\alpha$\,=\,0.24\%. The visibility improvement of 1.4\% with respect to the all-SMF configuration is in excellent agreement with our theoretical prediction, which also implies that the dispersion cancellation was complete. Lastly, Fig.~4(d) shows an interferometer with dispersion canceled nonlocally, corresponding to the case $\Phi_s$\,=\,$-\Phi_i$. In this configuration, the entire signal arm was made of SMF, with a differential second order dispersion of $\Delta (\beta_2 L)$\,=\,$-2.2\times 10^{-2}$\,ps$^2$. In the idler arm, the long path comprised 269.5\,cm of LEAF fiber and 18.0\,cm of SMF, whereas the short path used 190.0\,cm of SMF so that the differential dispersion came to $+2.2\times 10^{-2}$\,ps$^2$. For Fig.~4(d) configuration, we measured a raw visibility 99.6\,$\pm$\,0.2\% that is higher than the 98.2\% visibility obtained in the dispersion-limited case of Fig.~4(a). Together with the measurements for Figs.~4(b) and 4(c), we have clearly demonstrated nonlocal cancellation of dispersion in Franson interferometry. Additionally, for configurations 4(a), (c), and (d), we measured the Franson visibility at increasing pair generation rate. The results in Fig.~5 are in good agreement with the expected theoretical relationship $V\approx1-\alpha$ \cite{marc2}. At 99.6\,$\pm$\,0.2\% we have achieved near-unity Franson visibility that is limited only by the mean pair per gate $\alpha=0.24\%$ and the remaining accidental coincidences with experimental uncertainties. Note that successful cancellation is consistently maintained at each measurement with different $\alpha$.

\begin{figure}[hbt]
\includegraphics[width=0.48\textwidth]{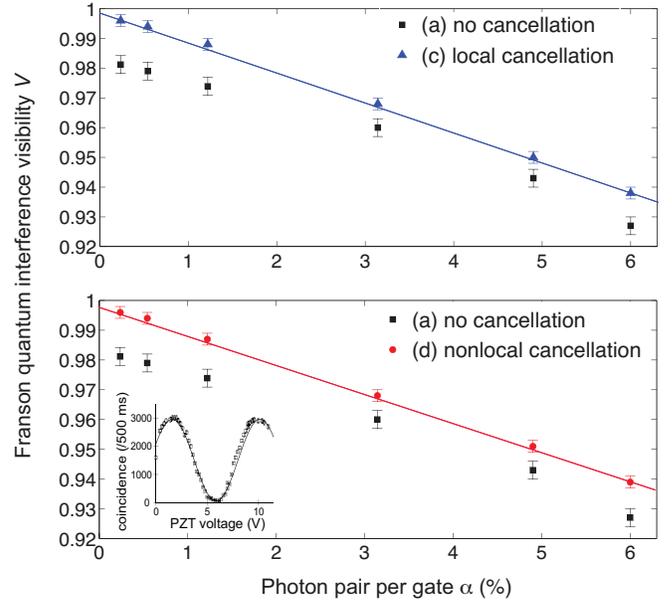}% Here is how to import EPS art
\caption{\label{f5} Measured raw Franson visibility versus generated pair per gate, for local (upper panel) and nonlocal (lower panel) cancellation configurations. Error bars are based on one standard deviation of 100 measurement samples each with an integration time of 0.1 s. The linear fits (solid lines) represent the $V=1-\alpha$ functional dependence. The inset shows the interference fringe of coincidences corresponding to 99.6\% visibility using nonlocal cancellation.}
\end{figure}

The raw visibility we measured via dispersion cancellation represents a significant violation of Bell's inequality by 145 standard deviations. We believe it is the highest violation ever reported for Franson interferometry, and for the first time the measured quality of time-energy entanglement is at the same level as that of its polarization counterpart \cite{kwiat, kim}. Maximal violation of CHSH inequality is essential in quantum information applications, including certified random number generation \cite{pironio}, remote state preparation, and quantum repeaters. An immediate application is high-dimensional time-energy entanglement-based quantum key distribution where multiple time bits are encoded per photon-pair coincidence \cite{Zhong}. Achieving near-unity Franson visibility can significantly improve the secure key rate by putting a tighter bound on Eavesdropper's accessible information \cite{hdQKD}. Another nonlocal cancellation application is on-chip Franson interferometric measurement in which a semiconductor waveguide-based MZI possesses a significant imbalance of dispersion between two highly unequal paths, and such a GVD mismatch is difficult to cancel locally and quickly. It is thus desirable to keep the on-chip MZI as one arm of the Franson interferometer, and apply nonlocal dispersion cancellation at the remote arm by implementing a fiber-based MZI whose dispersion can be easily engineered to cancel the on-chip dispersion. 

In summary, we have rigorously investigated the dispersive effect in nonlocal Franson interferometry. For the first time, we have demonstrated complete cancellation of femtosecond-level dispersion in a strictly nonlocal sense, and recovered a nearly perfect quantum interference visibility that represents the highest level of entanglement quality measured so far for time-energy entangled photons. Dispersion canceled Franson interferometry is relevant to a range of temporal measurements and applications exploiting entangled states of light.

The authors acknowledge valuable technical discussion with J.\ H.\ Shapiro. This work was supported by the DARPA InPho program, including US Army Research Office award W911NF-10-1-0416.


\begin{thebibliography}{20}%{apssamp} Produces the bibliography via BibTeX.
\bibitem{CHSH} J. F. Clauser, M. A. Horne, A. Shimony, and R. A. Holt, Phys. Rev. Lett. {\bf 23,} 880 (1969).

\bibitem{Franson1989} J. D. Franson, Phys. Rev. Lett. {\bf 62,} 2205 (1989).

\bibitem{kwiat} P. G. Kwiat, E. Waks, A. G. White, I. Appelbaum, and P. H. Eberhard, Phys. Rev. A, {\bf 60,} R773 (1999).

\bibitem{kim} F. N. C. Wong, J. H. Shapiro, and T. Kim, Laser Phys. {\bf 16,} 1517 (2006).

\bibitem{Zhong} T. Zhong, F. N. C. Wong, A. Restelli, and J. C. Bienfang, Opt. Express {\bf 20,} 26868 (2012).

\bibitem{Steinberg} A. M. Steinberg, P. G. Kwiat, and R. Y. Chiao, Phys. Rev. A {\bf 45,} 6659 (1992).

\bibitem{Giovannetti} V.Giovannetti, S. Lloyd, L. Maccone, and F. N. C. Wong, Phys. Rev. Lett. {\bf 87,} 117902 (2001).

\bibitem{qoct} M. B. Nasr, B. E. A. Saleh, A. V. Sergienko, and M. C. Teich, Phys. Rev. Lett. {\bf 91,} 083601 (2003).

\bibitem{Resch} R. Kaltenbaek, J. Lavoie, D. N. Biggerstaff, and K. J. Resch, Nat. Phys. {\bf 4}, 864 (2008).

\bibitem{pc-oct} J. Le Gou\"{e}t, D. Venkatraman, F. N. C. Wong, and J. H. Shapiro, Opt. Lett. {\bf 35,} 1001 (2010). 

\bibitem{Franson92} J. D. Franson, Phys. Rev. A {\bf 45,} 3126 (1992).

\bibitem{Donnell} K. A. O'Donnell, Phys. Rev. Lett. {\bf 106,} 063601 (2011).

\bibitem{Kim} S.-Y. Baek, Y.-W. Cho, and Y.-H. Kim, Opt. Express {\bf 17,} 19241 (2009).

\bibitem{Wasak} T. Wasak, P. Szankowski, W. Wasilewski and K. Banaszek, Phys. Rev. A {\bf 82,} 052120 (2010).

\bibitem{Franson10} J. D. Franson, Phys. Rev. A {\bf 81,} 023825 (2010). 

\bibitem{Shapiro} J. H. Shapiro, Phys. Rev. A {\bf 81,} 023824 (2010).

\bibitem{Fransonclass} J. D. Franson, Phys. Rev. Lett. {\bf 67,} 290 (1991).

\bibitem{epr} A Einstein, B. Podolsky, and N. Rosen, Phys. Rev. {\bf 47,} 777 (1935).

\bibitem{marc2} I. Marcikic, H. de Riedmatten, W. Tittel, V . Scarani, H. Zbinden, and N. Gisin, Phys. Rev. A {\bf 66,} 062308 (2002).

\bibitem{pironio} S. Pironio, A. Ac\'in, S. Massar, A. Boyer de la Giroday, D. N. Matsukevich, P. Maunz, S. Olmschenk, D. Hayes, L. Luo, T. A. Manning, and C. Monroe, Nature {\bf 464,} 1021 (2010).

\bibitem{hdQKD} T. Brougham, S. M. Barnett, K. T. McCusker, P. G. Kwiat, and D. J. Gauthier,  J. Phys. B: At. Mol. Opt. Phys. {\bf 46,} 104010 (2013).
\end{thebibliography}
\end{document}